\documentstyle[11pt,newpasp,twoside,epsf]{article}
\begin{document}
\title{The Properties of Young Clusters in M82}
 \author{Linda J. Smith}
\affil{Department of Physics and Astronomy, University College London,
Gower St., London WC1E 6BT, UK}
\author{John S. Gallagher III}
\affil{Department of Astronomy, University of Wisconsin-Madison,
5534~Sterling, 475 North Charter St., Madison WI 53706, USA}
\begin{abstract}
We present a detailed study of two luminous super star clusters in the
starburst galaxy M82. Spectra, covering 3250--8790\,\AA\ at a
resolution of 1.6\,\AA, were obtained at the 4.2\,m William Herschel
Telescope (WHT) for cluster F and the highly reddened cluster L.  We
compare the strengths of the observed Balmer absorption lines and the
Balmer jump in the blue spectrum of F with theoretical model cluster
spectra using the PEGASE spectral synthesis code to derive an age of
$60\pm20$\,Myr. For cluster L, we find that the similarities in the
strength of the Ca\,II triplet and overall spectral appearance with
cluster F suggest a similar age.  The brightness and compactness of
cluster F make it an ideal candidate for determining its dynamical
mass by measuring the velocity dispersion.  We present the results of
such an investigation based on echelle spectra at a resolution of
8\,km\,s$^{-1}$ obtained at the WHT from 5760--9140\,\AA.
By cross-correlating various wavelength
regions in the spectrum of cluster F with cool giant and supergiant
template stars, we derive a velocity dispersion and, by application of
the virial theorem, determine a dynamical mass of 2$\times
10^6$\,M$_\odot$.  We compare our derived mass with those determined
for other young super star clusters and discuss whether our derived
parameters are consistent with cluster F being able to survive to
become a globular cluster.
\end{abstract}


\section{Introduction}
Observations with the {\it Hubble Space Telescope (HST)} have revealed
that hundreds of super star clusters (SSCs) are present in the nearby
starburst galaxy M82 (O'Connell et al. 1995). SSCs appear to be
frequently associated with starbursts, and it has often been suggested
that they represent young globular clusters.
One critical aspect is whether they have enough mass to survive over long
time-scales. In this paper, we present a detailed investigation
of the luminous cluster F (O'Connell \& Mangano 1978) and the nearby,
highly reddened, cluster L (Kronberg, Pritchet, \& van den Bergh 1972).
They are located 440 pc south-west of the nucleus of M82.

Our study is based on optical spectroscopy
obtained at the 4.2\,m William Herschel Telescope (WHT) at resolutions
of 1.6\,\AA\ and 8\,km\,s$^{-1}$. We use the intermediate dispersion
spectra to obtain ages for the two clusters.
The brightness and compactness of
SSC F make it an ideal candidate for measuring the line of sight
velocity dispersion, and hence obtaining the dynamical mass of the
cluster. We use our recently obtained high dispersion red spectra to
derive the mass and discuss whether our derived parameters are
consistent with SSC F being able to survive to become a globular
cluster.

\section{The Ages of Clusters F and L}
Observations of clusters F and L were obtained in 1997 March with the
WHT on La Palma, Canary Islands, the double beam spectrograph ISIS and
$1024 \times 1024$ TeK CCDs on the blue and red arms. The spectra,
shown in Fig. 1, cover 3300--5500 \AA\ and 5700--8800 \AA\ for cluster F
and 5700--8800 \AA\ for cluster L. The resolution is
1.6 \AA\ with a S/N of 30--40 for a total exposure time
of 100 min for each arm. We derive a $V$ magnitude of 15.8 and
(B$-V$)=1.07 for cluster F.
\begin{figure}
\epsfbox[71 243 453 673]{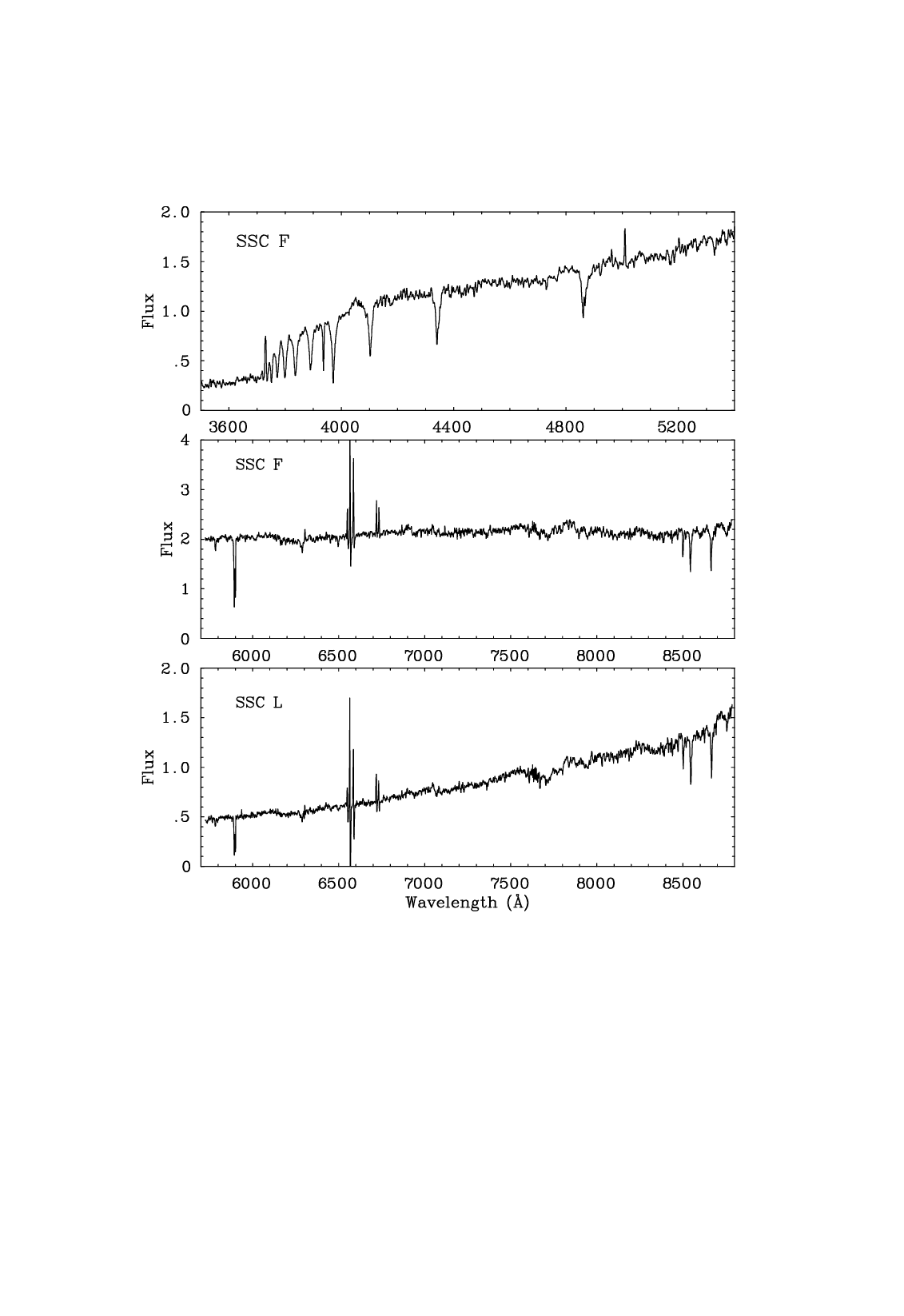}
\caption{The WHT$+$ ISIS spectra of SSCs F and L
(smoothed with $\sigma=1.0$\,\AA\ for presentation purposes).
The y-axis is in units of $10^{-15}$\,ergs\,s$^{-1}$\,cm$^{-2}$\,\AA$^{-1}$.}
\end{figure}

The blue spectrum of cluster F is dominated by broad Balmer absorption
lines and a strong Balmer jump, indicative of a mid-B spectral
type. The red spectrum shows strong absorption due to the Ca\,II
triplet and many weak features attributable to F and G supergiants.
The red spectrum of cluster L is very similar to that of F.
\begin{figure}
\epsfbox[81 276 464 512]{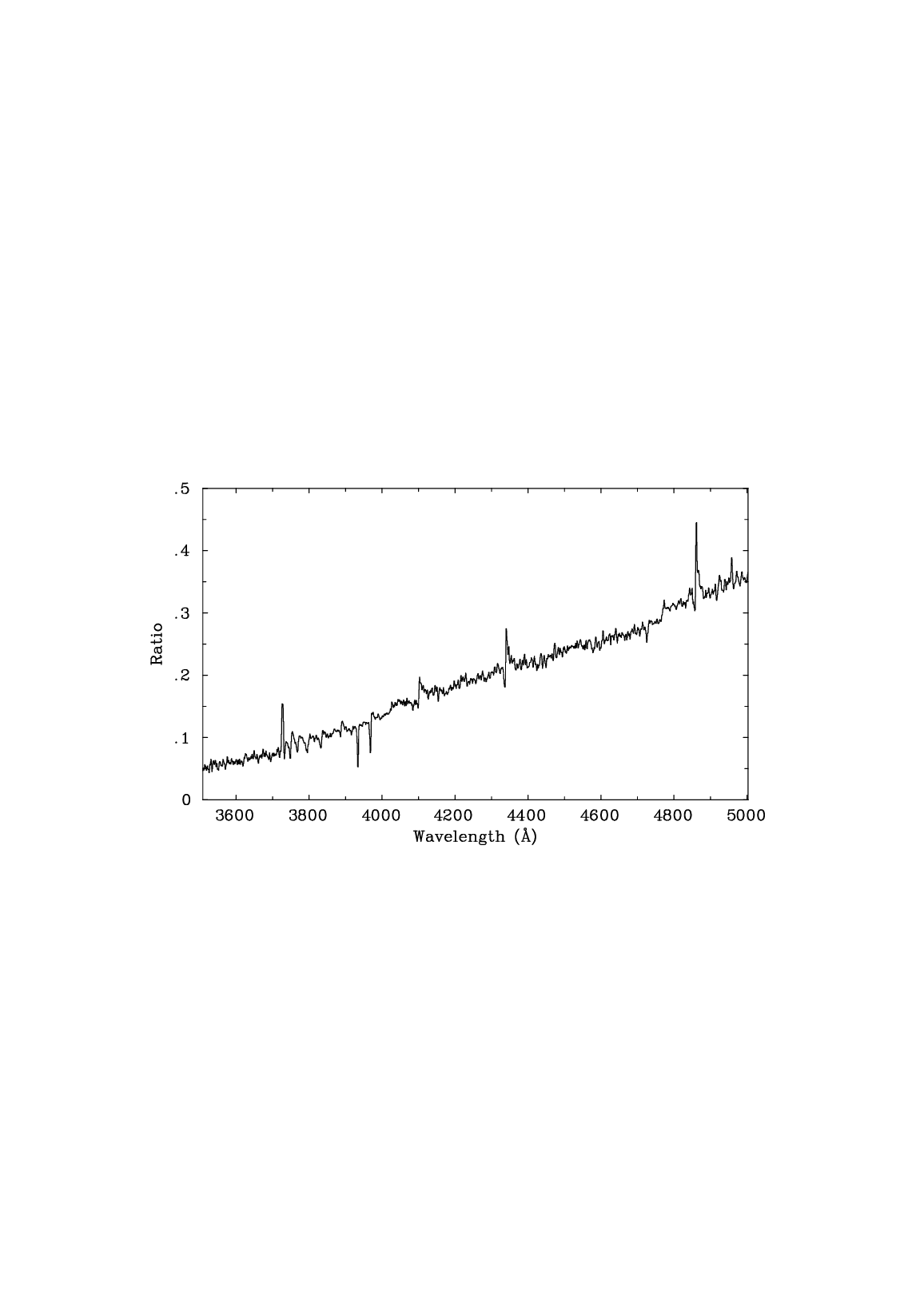}
\caption{The ratio between the observed cluster F spectrum and the 60 Myr
PEGASE model. The slope is caused by dust obscuration in M82.}
\end{figure}

The quality of the WHT spectra allows us to considerably improve
previous age estimates for cluster F. We have used the PEGASE spectral
synthesis code (Fioc \& Rocca-Volmerange 1997) to compute synthetic
spectra for ages of 20, 40, 60, 80 and 100 Myr, using a Salpeter IMF,
the Geneva evolutionary tracks and the Jacoby, Hunter, \& Christian
(1984) spectral library.
The ratio of the observed to model spectra suggests a reddening
$E(B-V) \approx 1.5$, although we find that a standard Galactic
extinction law does not properly correct the data.  The 40 Myr model
gives the best fit to the H$\beta$, H$\gamma$ and H$\delta$ absorption
lines but does not do as well for the Balmer jump region as the 60 and
80 Myr models. The 20 and 100 Myr models gave worse fits in all
areas. We therefore adopt an age for cluster F of $60\pm20$\,Myr.
The ratio between the observed spectrum of cluster F and the 60 Myr
model is shown in Fig. 2.

For cluster L, we find that similarities in the strength of the Ca\,II
triplet and the overall spectral appearance with cluster F suggest a
similar age. We conclude that M82 experienced an episode of intense
star formation $\approx$ 60 Myr ago in the mid-disc region. In the
central region, the age of the starburst is younger, at 20--30 Myr
(Rieke et al.  1993).  More details of this analysis can be found in
Gallagher \& Smith (1999).

\section{The Mass of Cluster F}
Cluster F was observed in 1999 February with the WHT and the Utrecht
echelle spectrograph (UES). The detector used was a $2048 \times 2048$ SITe
CCD and the wavelength range covered was 5760--9140 \AA\ in a single
exposure at a resolution of 8~km\,s$^{-1}$. We achieved a 
per pixel S/N of 15--25 in a total integration time of 6.4 hr.

We chose this spectral region because, as discussed by Ho \&
Filippenko (1996a,b), and demonstrated by our ISIS spectra, the
wavelength region longward of 5000 \AA\ is dominated by the light of
cool supergiants.  Thus by cross-correlating the cluster spectrum with
a suitable template spectrum of a cool supergiant, it is possible to
recover the velocity dispersion of the cluster, and hence derive its
mass by application of the virial theorem.

We obtained UES spectra of eight stars with spectral types from
A7\,III to M2\,Iab for the purpose of providing a suitable template.
Comparison of these spectra with that of cluster F shows that the
best spectral match is between HR 6863 (F8\,II) and HR 1529 (K1 III), as
demonstrated in Fig. 3 for the region containing the Ca\,II triplet.

\begin{figure}
\epsfbox[90 287 450 505]{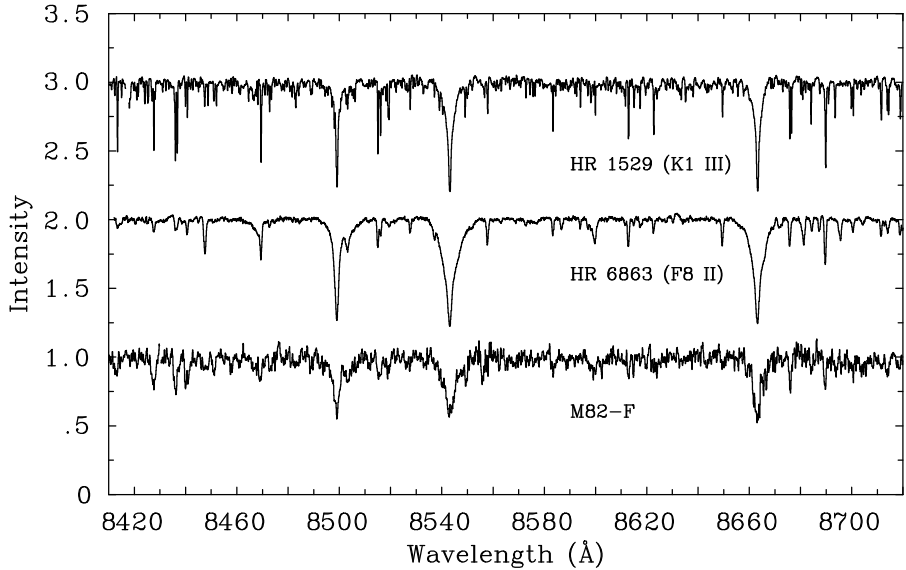}
\caption{WHT$+$UES spectra covering the Ca\,II triplet region at a resolution
of 8 km\,s$^{-1}$ for M82-F and the two template stars.
The data for M82-F have been lightly smoothed for
presentation purposes.}
\label{ues}
\end{figure}
We have therefore used these two stars as templates and
cross-correlated their spectra with that of cluster F. We chose four
spectral regions which are free of telluric absorption lines:
6010--6275 \AA, 6320--6530 \AA, 7340--7590 \AA, and 7705--8132 \AA\
for the analysis.  We ignored the region containing the Ca\,II
triplet because these lines are saturated in the template spectra and
broad Paschen absorption lines from early-type stars are present in
the cluster F spectrum.

For each cross-correlation function, we measured the FWHM by fitting
a Gaussian profile. The relationship between the FWHM and
the velocity dispersion was empirically calibrated by broadening the
template spectra with Gaussians of different dispersions and cross-correlating
with the original spectra.

Using this approach, we find that relative to HR 6863
(F8\,II) and HR 1529 (K1 III) the velocity dispersion of F is
$13.2\pm1.9$ and $16.5\pm2.4$~km\,s$^{-1}$ respectively. The latter
value is larger because the dispersion due to macroturbulence is
smaller in the K1 giant. The difference in the two values is consistent
with the velocity dispersion derived by cross-correlating the two
template stars. We adopt the lowest value as representing the
velocity dispersion of F because we expect the spectrum to be dominated
by cool supergiants, and the work of Gray \& Toner (1986, 1987) indicates
that the macroturbulence dispersion in an F8 bright giant is similar to
that of cool supergiants.

To derive the dynamical mass of cluster F, we need its half-light
radius.  O'Connell et al. (1995) derive a radius of 1.9\,pc from HST
images taken with WFPC. We have searched the HST archive for images of cluster
F taken with WFPC2 and have found two suitable 
images. We measure a half-light radius of 4.4\,pc, uncorrected
for the PSF which is likely to reduce it by $\sim$ 10\%.
We assume that this radius represents a good upper bound to
the half-mass radius, and that the cluster is
spherically symmetric with an isotropic velocity distribution.
Application of the virial theorem then gives a mass of $1.8\pm0.5 \times
10^6$ M$_\odot$. 
\begin{table}
\caption {Comparison of Properties of Super Star Clusters}
\begin{tabular}{llclcc}
\tableline
Cluster & Age & M$_V$ & R$_h$ & $\sigma$ & Mass \\
       & (Myr) & (mag) & (pc) & (km\,s$^{-1}$) & (M$_\odot$)\\
\tableline
NGC 1569-A$^{a,c}$ & 10--20 &$-14.1$ & $1.9\pm0.2$ & $15.7\pm1.5$ & 
$1.1\pm0.2 \times 10^6$ \\
NGC 1705-1$^{b,c}$ & 10--20 & $-14.0$ & $0.9\pm0.2$ & $11.4\pm1.5$ & 
$2.7\pm0.8 \times 10^5$ \\
M82-F & $60\pm20$ & $-15.8$ & $4.4$ & $13.2\pm1.9$ & $1.8\pm0.5 \times 10^6$\\
\tableline
\tableline
\noalign{$^a$Ho \& Filippenko (1996a); $^b$Ho \& Filippenko (1996b);
$^c$Sternberg (1998).}
\end{tabular}
\end{table}
\section{Is Cluster F a Young Globular Cluster?}
In Table 1, we compare our derived properties of cluster F with those
derived for the young super star clusters NGC 1569-A and NGC 1705-1
(Ho \& Filippenko 1996a,b; Sternberg 1998).  
Cluster F appears to be more massive and
luminous than both these clusters. To derive the value of M$_V$ given
in Table 1, we have measured a $V$ magnitude of 16.5 from the WFPC2
image, and assumed a reddening of $E(B-V)=1.5$ (Sect. 2).

Mandushev et al. (1991) provide measurements for 32 Galactic globular
clusters. The mass of cluster F is a factor of $\sim 9$ higher than
their mean mass of $1.9 \times 10^5$ M$_\odot$. From the PEGASE
models, we expect cluster F to dim by 4.5 mag for an age of 15 Gyr,
giving M$_V=-11.3$. This is higher than the mean value from Mandushev
et al. (1991) of $-8.1\pm1.2$ although the reddening towards F is
uncertain.  If we use $E(B-V)=1.0$ (O'Connell et al. 1995) instead,
then the predicted M$_V$ at 15 Gyr of $-9.8$ agrees better with the
mean value of Mandushev et al. (1991). The predicted mass-to-light
ratio $(M/L_V)_\odot$ for cluster F of 0.6 or 2.5 (with M$_V=-11.3$
and $-9.8$, respectively) agrees well with the range of 0.7--2.9 for
Galactic globular clusters (Mandushev et al. 1991).
This suggests that cluster F could survive to become a globular
cluster. Although it would be considerably more massive than the average 
Galactic globular cluster, it still would have less mass than large globular
clusters such as $\omega$ Cen (Merritt, Meylan, \& Mayor 1997).

\end{document}